\documentclass[aps,pra,reprint,a4paper,showpacs,amsmath,amssymb]{revtex4-1}
\pdfoutput=1
\usepackage{bm}
\usepackage{hyperref}
\usepackage{xcolor}
\usepackage{overpic}
\usepackage{mathtools}
\usepackage{bbold}
\usepackage{amsmath}
\usepackage{graphicx}
\graphicspath{{figures/}}
\newcommand{\dd}{\text{d}}
\newcommand{\ee}{\text{e}}
\newcommand{\ii}{\text{i}}

\renewcommand{\vec}[1]{\bm{#1}}
\begin{document}
\title{
Order-by-Disorder Degeneracy Lifting of Interacting Bosons on the Dice Lattice}
\author{Matja\v{z} Payrits}
\author{Ryan Barnett}
\affiliation{Department of Mathematics, Imperial College London,
London SW7 2AZ, United Kingdom}
\begin{abstract}
Motivated by recent experimental progress in the realization of synthetic gauge
fields in systems of ultracold atoms, we consider interacting bosons on the
dice lattice with half flux per plaquette.  All bands of the non-interacting
spectrum of this system were previously found to have the remarkable property
of being completely dispersionless.  We show that degeneracies remain when
interactions are treated at the level of mean field theory, and the ground
state exhibits vortex lattice configurations already established in the simpler
XY model in the same geometry.  We argue that including quantum and thermal
fluctuations will select a unique vortex lattice up to overall symmetries based
on the order-by-disorder mechanism.  We verify the stability of the selected
state by analyzing the condensate depletion.  The  latter is shown to exhibit
an unusual non-monotonic behavior as a function of the interaction parameters
which can be understood as a consequence of the dispersionless nature of the
non-interacting spectrum.  Finally, we comment on the role of domain walls
which have interactions mediated through fluctuations.
\end{abstract}
\maketitle

\section{Introduction}

The experimental realization of synthetic gauge fields in systems of ultracold
atoms by several groups has reinvigorated interest in lattice boson systems
under large effective magnetic fields (for a recent review, see
Ref.~\cite{GoldmanWow}).  Recently, the Hofstadter model, which describes
particles on a square lattice under an external effective magnetic field, was
realized in a regime where the magnetic length is on the order of the lattice
constant \cite{Aidelsburger2013,Miyake2013}.  For electrons in solid state
materials, such field strengths would correspond to extremely large magnetic
fields on the order of $10^4$ Tesla.  Additionally, more complex optical
lattices can be realized, including the Kagom\'{e} lattice \cite{Jo2012}, which
opens the door to exploring the generalizations of the Hofstadter model to
non-Bravais lattices.

Weakly interacting bosons at low temperature on a lattice under an external
effective magnetic field, created from rotation or a synthetic gauge field,
will generally form a vortex lattice \cite{Cooper2008, Fetter2009}.  When the
magnetic length is large compared to the lattice constant, the familiar
triangular Abrikosov vortex lattice is formed.  However, when the magnetic
length is on the order of the lattice constant, there can be a subtle interplay
between these two length scales and a variety of non-triangular lattices can be
formed as seen from the minimization of the XY model or the Gross-Pitaevskii
energy functional \cite{Teitel1983,Halsey1985,Straley1993,Powell2010}. Such
mean field theories are typically sufficient to determine the ground state of
these systems up to overall symmetries.

The dice lattice under an effective magnetic field is an exception to this
paradigm.  For the particular case of half a flux quantum  per plaquette, it is
known that its single-particle spectrum in the tight-binding approximation
exhibits three completely flat bands \cite{Vidal1998}. Interactions are
therefore necessarily of crucial importance in determining the possible phases
of this system.  Flat band systems with nonzero Chern numbers have also been
proposed as candidates for realizing fractional Chern insulators
\cite{Tang2011,Sun2011,Neupert2011}.  Though the Chern numbers of the bands in
the present analysis of the dice lattice vanish, it was found in
Ref.~\cite{Wang2011} that it is possible to obtain bands with non-zero Chern number
through the introduction of spin-orbit coupling (though this introduces
dispersion to the bands). Additionally, flat band systems have received
considerable attention within the context of ultracold gases
\cite{Bergman2008,Huber2010,Baur2012,Jo2012,You2012,Li2013,Yao2013}.  An
analysis by Korshunov of the related XY model on the dice lattice with half
flux per plaquette shows that such degeneracies persist in the vortex lattice
structures \cite{Korshunov2001,Korshunov2004, Korshunov2005}. In particular, it
was found that the four periodic vortex lattice structures shown in
Fig.~\ref{fig:korsh} are all degenerate ground states of the XY model.

The dice lattice first garnered considerable attention in the context of the
so-called \emph{topological} or \emph{Aharonov-Bohm} localization mechanism
\cite{Sutherland1986,Vidal1998,Morita1972}. Later studies have
explored  the phase diagram of the Bose-Hubbard model in a dice
geometry in various approximations \cite{Cataudella2003,Rizzi2006},
and shown that it can give rise to effective Dirac-Weyl fermions
\cite{Bercioux2009}.  In Ref.~\cite{Burkov2006} it was shown that an
order-by-disorder mechanism in the dice lattice Bose-Hubbard model yields a
Vortex-Peierls state near the Mott insulating - superfluid transition.  Most
recently, ground state quantum phases of the model were determined in the
lowest Landau level regime where it is appropriate to project into the lowest
single-particle band of the system~\cite{Moller2012}. Some of the exotic
properties associated with the single-particle spectrum of the dice lattice
have also been observed experimentally \cite{Abilio1999,Serret2002,Tesei2006}.

In this work, we consider interacting bosons with nearest-neighbor hopping on
the dice lattice with half flux per plaquette, described by the Bose-Hubbard
model.  We show that the same periodic vortex lattices as established for the
XY model~\cite{Korshunov2004} are degenerate energetic minima at the level of
Gross Pitaevskii mean field theory.  As these degeneracies are not protected by
any symmetry, they are expected to be lifted by the order-by-disorder mechanism
\cite{Villain1980,Henley1989}. Although quantum order-by-disorder is perhaps
most familiar from frustrated magnetism, it  can also play important roles in
ultracold atomic gases \cite{Song2007,Turner2007,Zhao2008,You2012,Barnett2012}.
We in particular show that quantum and thermal fluctuations at quadratic order
completely lift the degeneracy between the four candidate vortex lattices shown
in Fig.~\ref{fig:korsh}, where state (b) acquires the lowest energy.  In
contrast, it was found that thermal fluctuations in the analogous classical XY
model on the dice lattice at quadratic order do not lift the degeneracy between
the ground state vortex configurations and one must rely on anharmonic
fluctuations which are estimated to be small on experimental scales
\cite{Korshunov2005}. It should be noted that Ref.~\cite{Burkov2006}
performed a similar analysis near the Mott insulator-superfluid transition,
whereas we focus on the deep superfluid regime.

This paper is organized as follows.  In Sec.~\ref{sec:setup}, we set up the
theoretical problem of bosons hopping on the dice lattice under an effective
magnetic field.  For completeness, we also review the single-particle spectrum
of flat bands originally found in Ref.~\cite{Vidal1998}.  In Sec.~\ref{sec:mft}
we determine the periodic mean field vortex lattice configurations.  These are
shown to have the same phase structure as reported on studies of the XY model
\cite{Korshunov2001}.  Unlike the  XY model, the GP energy functional has local
density degrees of freedom, which are shown to take on two distinct values
representing the two inequivalent sites in the dice lattice.  In
Sec.~\ref{sec:obd} we describe the computation of the collective excitations
about each of the four periodic vortex configurations.  We will show that
harmonic fluctuations completely lift the degeneracy between the configurations
through the order-by-disorder mechanism.  We address the stability of the
proposed state with respect to quantum and thermal depletion.  As is common in
two-dimensional systems \cite{Mermin1966}, the thermal depletion exhibits a
logarithmic infrared divergence which is removed for finite-sized systems.  We
show that the depletion is small for realistic experimental parameters.
Interestingly, the depletion shows a non-monotonic behavior as a function of
the interaction parameters which can be attributed to the flat band structure.
Finally, in Sec.~\ref{sec:discussion} the results are discussed and the work is
concluded.

\section{Theoretical setup}
\label{sec:setup}

\begin{figure}
\includegraphics{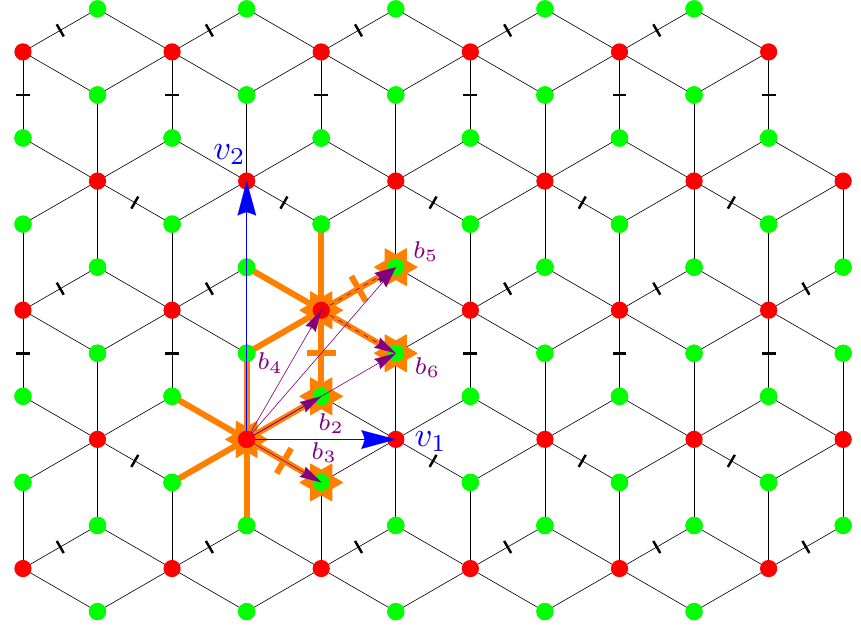}
\caption{(Color online) The dice lattice under an effective magnetic field using the
gauge of Ref.~\onlinecite{Moller2012}. There are two types of sites: hub sites
with a coordination number of 6 and rim sites with a coordination number of 3.
The links and sites outlined in orange comprise the half-flux-per-plaquette
magnetic unit cell. Particles  acquire a phase of $\pi$ when hopping across
crossed links and no phase when hopping across uncrossed links.  The lattice
vectors $\vec{v}_1$ and $\vec{v}_2$ can be chosen to be orthogonal, as in the
figure where $\vec{v}_1=\left( 1,0 \right)$, $\vec{v}_2=\left( 0,\sqrt{3}
\right)$.  For convenience we have set the lattice constant to unity. The
lattice can thus be viewed as a rectangular Bravais lattice with a 6-fold
basis. The basis vectors are $\vec{b}_1=\vec{0},\vec{b}_2=\left(
1/2,1/2\sqrt{3} \right),\vec{b}_3=\left( 1/2,-1/2\sqrt{3} \right),
\vec{b}_4=\left( 1/2,\sqrt{3}/2 \right),\vec{b}_5 = \vec{b}_2+\vec{b}_4$ and
$\vec{b}_6=\vec{b}_3+\vec{b}_4$.  In the absence of a gauge field the smallest
unit cell is half the size and we can take $\vec{v}_1$ and $\vec{b}_4$ for the
lattice vectors and $\vec{b}_{1-3}$ for the three basis vectors.}
\label{fig:simple-magnetic-dice-lattice}
\end{figure}

We consider bosons in the dice lattice potential, also referred to as
the $\tau_3$ lattice \cite{Vidal1998}, shown in
Fig.~\ref{fig:simple-magnetic-dice-lattice}.  The bosons are treated
within the tight-binding approximation with nearest-neighbor hopping.
Experimental proposals to engineer the dice optical lattice under an
effective magnetic field are provided in
\cite{Moller2012,Burkov2006,Bercioux2009,Rizzi2006}.  Josephson
junction arrays with the appropriate geometry provide another
physically feasible avenue of realizing the dice lattice
experimentally.

We label the unit cells of the lattice with an integer $n$ and denote the
corresponding unit cell location by $\vec{R}_n$. When the
unit-cell/basis-vector decomposition is important, we label the site displaced
from the origin of the $n$-th unit cell by the basis vector ${\bf b}_\gamma$
with the pair $\left( n,\gamma \right)$. Otherwise we label each site by a
single integer $i$ and denote its location by $\vec{r}_i$.  We consider the
system subjected to a synthetic gauge field with vector potential $\vec{A}$
such that its line integral around any lattice plaquette equals $\pi$.  We can
draw clear analogies with electromagnetism: by considering the example of a
particle of charge $q$ in the presence of an electromagnetic potential
$\vec{A}_{\text{EM}}$, $\vec{A}$ is found to be analogous to
$\frac{2\pi}{\Phi_0}\vec{A}_{\text{EM}}$ where $\Phi_0=h/q$ is the charge-$q$
elementary magnetic flux quantum.  Likewise,
$\int_\mathcal{C}\vec{A}\cdot\dd\vec{r}$, with $\mathcal{C}$ a cyclic path
along the edges of a plaquette, is analogous to the charged particle's
corresponding Aharonov-Bohm phase. Since the latter equals $\pi$ when the
plaquette is threaded by half an elementary magnetic flux, our synthetic gauge
field configuration is analogously referred to as
half(-elementary)-flux-per-plaquette.

In the continuum, electromagnetism is introduced into the Hamiltonian via
minimal coupling, i.e.\ $\hat{\vec{p}}\rightarrow\hat{\vec{p}}-q\hat{\vec{A}}$.
The tight-binding equivalent of this procedure is Peierls substitution, i.e.\
substituting pairs of creation and annihilation operators according to
$\hat{a}_i^\dagger \hat{a}_j \rightarrow \ee^{\ii A_{i j}}\hat{a}_{i}^\dagger
\hat{a}_{j}$, where  $\hat{a}_i$ is the annihilation  operator for the  $i$-th
site and $A_{i j}=\int_{\vec{r}_j}^{\vec{r}_i}\vec{A}( \vec{r})\cdot\dd\vec{r}$
is the line integral of the vector potential between sites $j$ and $i$ involved
in the hopping. The noninteracting part of the Hamiltonian is thus $\hat{H}_0 =
-t \sum_{\langle i j\rangle}\big( \ee^{\ii A_{i j}}\hat{a}_i^\dagger \hat{a}_j
+ \text{h.c.} \big)$ where the sum is taken over all pairs of nearest neighbor
sites.

We assume the inter-site interactions are negligibly weak. The interaction part
of the Hamiltonian thus consists of terms of the form $\frac{1}{2}U_i
\hat{a}_i^\dagger\hat{a}_i^\dagger\hat{a}_i\hat{a}_i$,  where $U_i$ is the
positive onsite interaction.  It is evident from
Fig.~\ref{fig:simple-magnetic-dice-lattice} that there are two qualitatively
different types of sites with coordination numbers 6 and 3. Following
Ref.~\cite{Sutherland1986}, we  refer to these as \emph{hub} (*) sites and
\emph{rim} ($\Delta$) sites, respectively.  We accordingly regard $U_*$ and
$U_\Delta$ as independent parameters.  Introducing the chemical potential
$\mu$, the full Bose-Hubbard model on the dice lattice  reads
\begin{eqnarray}
	\hat{H} = -t\sum_{\langle i j\rangle}\bigg(&&\ee^{\ii A_{i
	j}}\hat{a}_i^\dagger \hat{a}_j + \text{H.c.} \bigg)\nonumber\\
	{}+&&\;\sum_i\left(
	\frac{1}{2}U_i\hat{a}_i^\dagger\hat{a}_i^\dagger\hat{a}_i\hat{a}_i
	-\mu\hat{a}_i^\dagger\hat{a}_i\right)
	\label{eq:full-hamiltonian}.
\end{eqnarray}

\subsection{Single-particle spectrum}
\label{subsec:single-particle-spectrum}

For completeness, we review the single particle spectrum of the dice lattice at
half-flux per plaquette which was  first derived in Ref.~\cite{Vidal1998}.  At
the non-interacting level this system already displays some remarkable
features.  To calculate the  spectrum we use the momentum space creation and
annihilation operators defined as
$
\hat{a}_{\vec{k}\gamma} = \frac{1}{\sqrt{N}} \sum_{n}\hat{a}_{n\gamma}\ee^{-\ii
	\vec{k}\cdot\vec{R}_n}
$
where $N$ is the number of unit cells in the system.  We adopt the convenient
gauge of Ref.~\cite{Moller2012},  shown in
Fig.~\ref{fig:simple-magnetic-dice-lattice}, where the effective hopping
parameters are real.  Owing to the periodicity of the lattice we can rewrite
the non-interacting portion of the Hamiltonian in
Eq.~(\ref{eq:full-hamiltonian}) as
\begin{equation}
	\hat{H}_0= \sum_{\vec{k}} \vec{\hat{a}}_{\vec{k}}^\dagger
	H_0({\vec{k}}) \vec{\hat{a}}_{\vec{k}}
  \label{eq:momentum-space-Hamiltonian}
\end{equation}
where $\vec{\hat{a}}_{\vec{k}}=\left[ \hat{a}_{\vec{k}1}, \hat{a}_{\vec{k}2},
\cdots, \hat{a}_{\vec{k}6} \right]^\top$\llap{,\hspace{2pt}} $H_0({\vec{k}})$
is a 6-by-6 Hermitian matrix, and the summation is over the first Brillouin
zone.   By inserting a full set of eigenvectors $\sum_{\gamma=1}^6
\vec{u}_{\vec{k}\gamma} \vec{u}_{\vec{k}\gamma}^\dagger$ on both sides of $H_0(
\vec{k})$ in Eq.~(\ref{eq:momentum-space-Hamiltonian}) we can express the
Hamiltonian in terms of new quasiparticle operators
$\hat{\alpha}_{\vec{k}\gamma}=\vec{u}_{\vec{k}\gamma}^\dagger
\vec{\hat{a}_{\vec{k}}}$ and their corresponding eigenvalues
$\lambda_{\vec{k}\gamma}$ as
\begin{equation}
	\hat{H}_0= \sum_{\vec{k}\gamma} \lambda_{\vec{k}\gamma}
	\hat{\alpha}_{\vec{k}\gamma}^\dagger \hat{\alpha}_{\vec{k}\gamma}.
  \label{eq:nice-momentum-hamiltonian}
\end{equation}
The fascinating outcome is that the energies $\lambda_{\vec{k} \gamma}$ have no
dispersion and remain constant throughout the Brillouin zone. There are three
doubly degenerate bands with $\lambda_{\vec{k}\gamma}=\pm \sqrt{6}t, 0$. For
the lowest and highest bands this follows from the fact that their states can
be expressed as a sum of completely localized eigenstates.

The Wannier functions, obtained from the Fourier transform of these extended
Bloch wave functions, provide a particularly convenient basis for describing
the single-particle states.  For the highest and lowest energy bands they are
both eigenstates of the non-interacting Hamiltonian and completely localized.
For both of these bands they span a hub site and its six surrounding rim sites.
The amplitude on the hub is $1/\sqrt{2}$ and $1/\sqrt{12}$ on the rim sites.
The  phase of rim site $j$ relative to the central hub $i$  is, in the gauge of
Fig.~\ref{fig:simple-magnetic-dice-lattice},  simply $A_{j i}$ in the lowest
band and $\pi-A_{j i}$ in the top band, i.e.\ either $0$ or $\pi$ in both
cases.  The existence of these localized states does not fall under any of the
disorder-based localization paradigms, such as Anderson localization
\cite{Anderson1958}, but follows solely from destructive interference within
the so-called Aharonov-Bohm cages on the lattice \cite{Vidal1998}.  The Wannier
functions corresponding to the zero-energy eigenstates, on the other hand, are
only exponentially localized,  so this simple explanation of flatness is not
applicable for this case.

\section{Mean field theory}
\label{sec:mft}
We proceed by reintroducing interactions and finding the ground states at the
mean-field level by solving the Gross-Pitaevskii equation. This is equivalent
to assuming that the wave function can be written as a tensor product of
independent coherent states for each site.  Accordingly, we can replace
operators with c-numbers
\begin{equation}
  \hat{a}_{i}\rightarrow a_{i}=\sqrt{n_{i}}\,\ee^{\ii \theta_{i}}
  \label{eq:hydrodynamic-operator-transformation}
\end{equation}
with similar expressions for momentum-space quantities. Here $n$ and $\theta$
are the density and phase variables. We will later be able to find elementary
excitations about these states  by means of Bogoliubov theory.

Given the simple structure of the single-particle spectrum, it is reasonable to
ask whether there exist any states that simultaneously minimize \emph{both} the
single-particle and the interaction part of the mean-field energy.  The former
is true when the state can be constructed as a linear combination of states in
the lowest single-particle band and the latter when it gives rise to uniform
densities $n_*$ and $n_\Delta$ on the hub and rim sublattices, respectively.
By writing the state as a linear combination of lowest-band eigenstates, one
finds that such uniform densities can only be obtained when $U_\Delta/U_*=2$.
For future reference we term this parameter configuration the special point.
Besides uniform densities, the state also has a simple phase picture. In
particular, we only encounter three distinct magnitudes of \emph{gauge
invariant phase differences}. These are defined as 
\begin{equation}
	\Phi_{i j}=\theta_i-\theta_j-A_{i j}
\end{equation}
and are indeed independent of our chosen gauge. We derive their values in the
next section.

We conjecture that the states globally minimizing the total mean field energy
away from the special point retain uniform densities on both sublattices.
This is motivated by  the fact that the proposed states
merge with what are provably the only global minima at the special point
and by our failing to find a physically reasonable mechanism capable of
breaking the density symmetry. In the following section we show that the
necessary condition of the states remaining local energy minima is satisfied.
At the uniform sublattice density configurations we can furthermore follow
\cite{Korshunov2005} to show that the phase profiles minimizing the energy are
identical to those at the special point.

\subsection{Mean field calculation of sublattice densities}
\label{subsec:hydro}
The mean field energy of the Hamiltonian~(\ref{eq:full-hamiltonian}) is
\begin{equation}
  E=-2 t\sum_{\left<i j\right>}\sqrt{n_i n_j}\cos\left(\Phi_{i
  j}\right)+\frac{1}{2}\sum_i U_i n_i^2-\mu \sum_i n_i
  \label{eq:hydrodynamic-hamiltonian}
\end{equation}
We will derive the equations of motion with the corresponding Lagrangian
\begin{equation}
  L=\sum_i \left( -n_i \dot{\theta}_i \right)-E.
  \label{eq:lamo-lagrangian}
\end{equation}
Expressed in terms of gauge invariant quantities, the Euler-Lagrange
equations read
\begin{eqnarray}
    \dot{n_i}&=&2t\sum_{j\in\mathcal{N}_i}\sqrt{n_i n_j}\sin\Phi_{i
    j}\label{eq:mfbro-minus}\\
    \dot{\Phi}_{i
    j}&=&t\sum_{i'\in\mathcal{N}_i}\sqrt{\frac{n_{i'}}{n_i}}\cos\Phi_{i i'} -
    t\sum_{j'\in\mathcal{N}_j}\sqrt{\frac{n_{j'}}{n_j}}\cos\Phi_{j
    j'}\nonumber\\
    {}&+&U_j n_j- U_i n_i.
  \label{eq:euler-lagrange}
\end{eqnarray}
In this expression, $\mathcal{N}_i$ denotes the  set of all sites neighboring
site $i$.
For the ground state we demand that the time derivatives on the LHS be zero.

We now insert the key assumption of uniform sublattice densities. Taking into
account the overall geometry, the second equation yields
\begin{equation}
  U_* n_*-U_\Delta n_\Delta = t\sqrt{\frac{n_\Delta}{n_*}}\sum_{i'\in
	  \mathcal{N}_*}\text{cos}\Phi_{* i'}-t
	  \sqrt{\frac{n_*}{n_\Delta}}\sum_{j'\in
		  \mathcal{N}_\Delta}\text{cos}\Phi_{\Delta j}.
  \label{eq:almost-n-equation}
\end{equation}
As remarked before, the phase profiles occurring at the special point still
solve the equations. Let us denote the three distinct phase difference
magnitudes comprising them by $\Phi_l>\Phi_m>\Phi_s>0$ ($l, m, s$ for large,
medium, small).  Since the factor $\sqrt{n_i n_j}$ equals $\sqrt{n_\Delta n_*}$
for any neighboring $i$ and $j$, we can rewrite
equation~(\ref{eq:mfbro-minus}), the continuity equation, as
$\text{sin}\,\Phi_l=\text{sin}\,\Phi_m+\text{sin}\,\Phi_s$.  The condition that
the sum of phase differences around a plaquette equal $\pm \pi$ imposes the
restrictions $2\Phi_s+2\Phi_l=\pi$ and $-\Phi_s+2\Phi_m+\Phi_l=\pi$. This
system of equations  yields $\Phi_s \approx 9.74^\circ, \Phi_m \approx
54.74^\circ$ and $\Phi_l \approx 80.26^\circ$, along with the useful identity
\begin{equation}
	\ee^{\ii \Phi_s} + \ee^{\ii \Phi_m} + \ee^{-\ii \Phi_l} = \sqrt{3}.
	\label{eq:useful-identity}
\end{equation}
While it can easily be seen that each of these three phase differences  appears
exactly once for links surrounding any rim site,  it can also be shown that
each appears exactly twice among the links surrounding any hub site, though the
procedure is tedious.

This phase configuration is identical to the one obtained by Korshunov for the
dice lattice XY model \cite{Korshunov2001}. This is so because the form of
Eq.~(\ref{eq:mfbro-minus}) is the same in both cases, as it does not depend on
the local interaction terms of the Hamiltonian. Furthermore, the factor
$\sqrt{n_i n_j}$ in Eq.~(\ref{eq:mfbro-minus}) is  constant for all pairs of
neighboring sites in both cases. It can thus be factored out when considering
the ground state. In  Ref.~\cite{Korshunov2001} this is due to  the author's
explicitly taking a uniform density across all sites, while in our case it is
due to the alternating nature of the uniform density hub and rim sublattices. 

We can in fact easily determine the sublattice density values. Taking the
features of the phase configuration and Eq.~(\ref{eq:useful-identity}) into
account, Eq.~(\ref{eq:almost-n-equation}) simplifies to
\begin{equation}
  U_* n_*-U_\Delta n_\Delta=\sqrt{3}t\frac{2 n_\Delta-n_*}{\sqrt{n_\Delta
  n_*}}.
  \label{eq:n-equation-pedestrian}
\end{equation}
Given the two interaction strengths, this equation can be solved to determine
the ratio of densities on the hub and rim sites, $n_*/n_\Delta$.  Note that at
the special point, where $U_\Delta = 2U_*$, one has the simplest case
$n_*=2n_\Delta$, as expected.  Finally, with this solution the chemical
potential is found to be
\begin{align}
	\mu = U_*n_* -2 t \sqrt{ \frac{3 n_\Delta}{n_*}} = U_\Delta n_\Delta -
	t \sqrt{ \frac{3 n_*}{n_\Delta}}
\label{eq:mu}
\end{align}

\subsection{Mean field periodic ground states}
\label{subsec:ground-states}

We can assign to each plaquette a vorticity of either $\pi$ or $-\pi$.  Through
a qualitative comparison of this vortex lattice with the two-dimensional
Coulomb gas, relevant since neutral superfluid vortices are known to have
approximately a logarithmic interaction, one can argue that the most
energetically favorable configuration will have each vortex surrounded by as
many neighbors of the opposite vorticity as possible. The vortices are pinned
to the sites of the dual lattice which in this case  is the Kagom\'{e} lattice.
The geometric frustration of the Kagom\'{e} lattice prevents the possibility of
a purely local prescription for the distribution of vortices  minimizing the
energy.  The vortex configuration of mean field ground states is demonstrably
composed of \emph{chains} of like-vortices of length three \footnote{With the
exception of triangular clusters of like-vortices around rim sites which could
be regarded as cyclic chains of length 3. These cannot occur.}.

\begin{figure*}
	\includegraphics{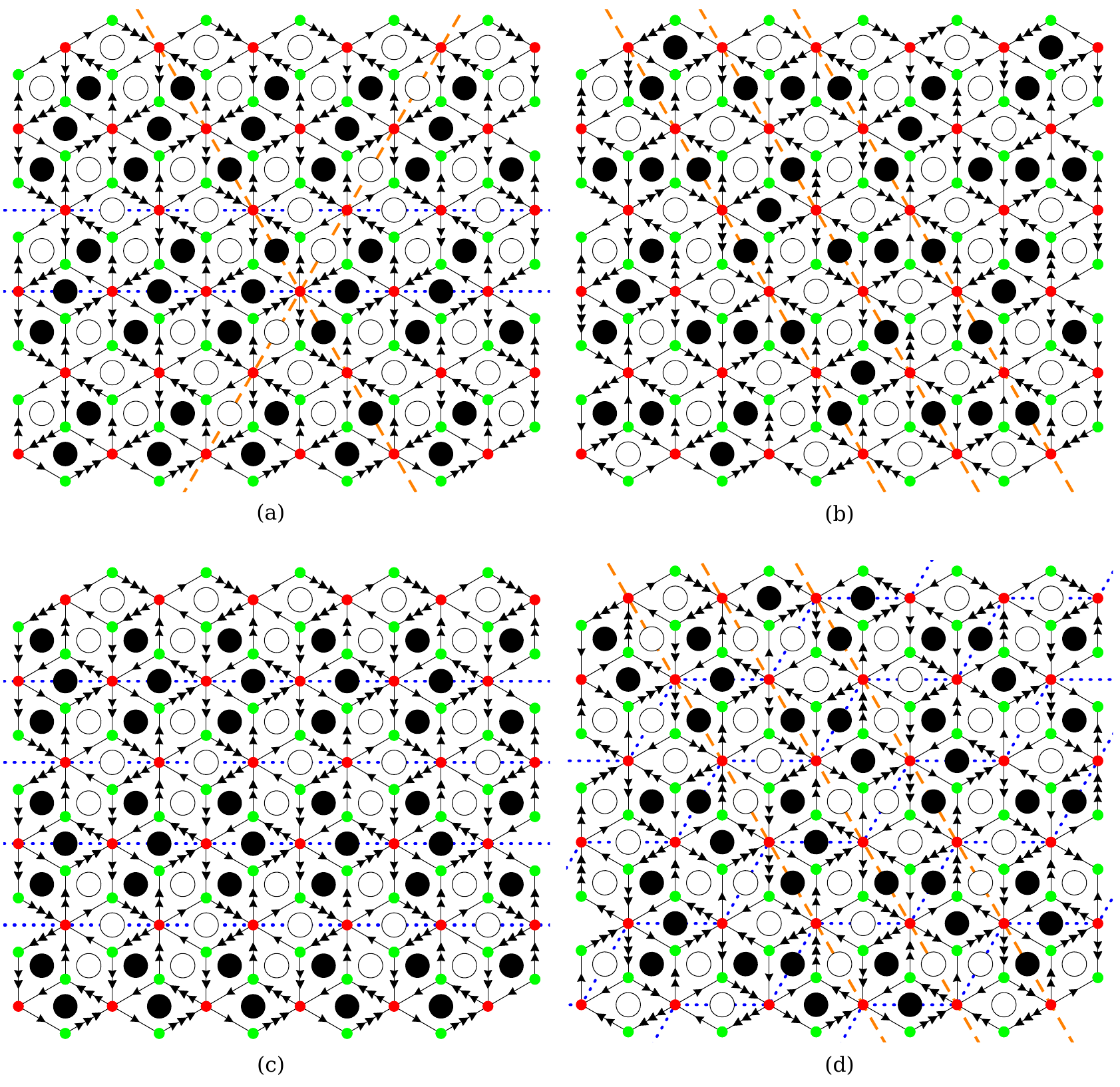}
\caption{(Color online) The four small unit cell periodic mean field ground states. The
single, double and triple arrows represent gauge invariant phase differences
$\Phi_s, \Phi_m$ and $\Phi_l$ across links, respectively, and the black (white)
disks represent positive (negative) plaquette vorticities. The dashed and
dotted lines signify locations of possible domain wall insertions (figure (a))
or domain walls themselves (all other figures). The dashed orange lines
represent type I domain walls while the blue dotted lines represent type II
domain walls.}
\label{fig:korsh}
\end{figure*}

Perhaps the simplest such state is shown in Fig.~\ref{fig:korsh}(a).  We can
obtain all other applicable states with only the three gauge invariant phase
differences introduced above by rearranging the phase differences along a
variety of infinite sequences of plaquettes in which every pair of neighboring
plaquettes shares just a single vertex.  We can think of this process as the
insertion of two types of \emph{zero-energy domain walls} into state~(a).  We
will refer to the domain walls we can insert parallel to the dashed lines  in
Fig.~\ref{fig:korsh}(a) as type I domain walls \footnote{We can in fact only
insert one of the type~I domain walls shown and all walls parallel to it, but
we cannot insert any two type I domain walls at an angle.}  and the ones we can
insert parallel to the dotted lines as type II domain walls. Inserting a type I
domain wall splits the lattice into two regions with orientations of the vortex
triads not parallel to the wall differing by $60^{\circ}$. A type II domain
wall bends the triads it crosses and establishes a mirror symmetry between both
of its sides. Type II domain walls also bend by $60^{\circ}$ whenever they
cross a type I domain wall \footnote{Further details of the phase permutations
comprising each type of domain wall and figures of single domain walls inserted
into state (a) are given in  Ref.~\cite{Korshunov2001}.}.

The unit cell of vortex state (a) contains six lattice sites. It is
twelvefold degenerate under the following geometric transformations that
preserve the Hamiltonian but not the state: translations by $\pm \vec{b}_4$ 
or $\vec{b}_4 -\vec{b}_1$, using the notation of
Fig.~\ref{fig:simple-magnetic-dice-lattice}, contributing a factor 2 to the
geometric degeneracy, the combination of time (arrow) reversal and spatial
inversion, contributing another factor of 2, and $\pm 2\pi/3$ rotations about
any site, contributing the final factor of 3. By inserting all possible type II
domain walls into (a) we obtain another twelvefold degenerate state with six
sites per unit cell, shown in Fig.~\ref{fig:korsh}(c), not related to state (a)
by geometric symmetries. Inserting all possible type I domain walls into state
(a) similarly yields the state shown in Fig.~\ref{fig:korsh}(b) with twelve
sites per unit cell.  Further inserting all possible type II domain walls into
(b) yields the state shown in Fig.~\ref{fig:korsh}(d), also containing twelve
sites per unit cell.  States (b) and (d) have a fourfold translational
degeneracy, so their total geometric degeneracy is 24-fold. Taking geometric
multiplicities into account this yields a total of 72 small unit cell mean
field periodic states, or SMPS's.  All other uniform sublattice periodic mean
field ground states can be obtained by gluing together the unit cells of the
above four classes of SMPS's \cite{Korshunov2005}.

It should be noted that given two asymptotically domain wall-free
regions, such that, for instance, the vortex lattice
is one of the 72 SMPS's on the far left and a distinct SMPS
on the far right, it is not in general possible to consistently interpolate
between the two through a sequence of SMPS regions, i.e.~state (a-d)-like
regions, glued by zero-energy domain walls. This implies either the possibility
of massive, i.e.~energetically costly, domain walls and point defects, or
global instabilities of such asymptotic configurations. We were unable to
obtain concrete results regarding this issue, but include a brief speculative
discussion of possible scenarios in the conclusion.

The geometric degeneracies discussed above originate from true symmetries of
the Hamiltonian and are as such not expected to be lifted by fluctuations. In
the following we focus on the effects of quantum zero-point and thermal
fluctuations on the four classes of accidentally degenerate periodic mean field
states (a-d) of Fig.~\ref{fig:korsh}. Later we also briefly comment on the
expected role of domain wall interactions in more general states.

\section{Order-by-disorder degeneracy lifting}
\label{sec:obd}

Through the mechanism of order-by-disorder~\cite{Villain1980}, quantum and
thermal fluctuations act to remove accidental mean-field degeneracies, i.e.\
degeneracies not protected by symmetries of the Hamiltonian.  In quantitative
terms, the state whose fluctuations yield the lowest Helmholtz free energy
$F=-\frac{1}{\beta}\ln{Z}$, where $Z={\rm Tr}(e^{-\beta \hat{H}})$ is the
partition function and $\beta=1/k_BT$, is selected.  At the level of Bogoliubov
theory, the excitation spectrum is described by independent harmonic
oscillators, and so we have
\begin{eqnarray}
	Z=\sum_{n_i}\ee^{-\beta \hbar \sum_j \omega_j\left( n_j+\frac{1}{2}
	\right)}= \prod_j \ee^{-\frac{\beta E_j}{2}}\frac{1}{1-\ee^{-\beta
	E_j}}.
	\label{eq:partition-function}
\end{eqnarray}
From this the free energy can be found to be
\begin{equation}
	F=\frac{1}{2}\sum_j E_j + \beta^{-1}\sum_j \ln\left( 1-\ee^{-\beta E_j}
	\right).
	\label{eq:free-energy}
\end{equation}
The first term corresponds to the zero-point quantum contribution to the free
energy and the second term to the contribution of thermal fluctuations. In our
case the index $j$ in Eq.~(\ref{eq:free-energy}) is a label for momentum and
band index.

\subsection{Collective Excitation Spectrum}

We now derive the collective excitation spectrum of
Eq.~(\ref{eq:full-hamiltonian}) at the level of Bogoliubov theory.  This
involves expressing the annihilation operators as $\hat{a}_i=a_i+\delta
\hat{a}_i$, where $a_i$ are the mean-field c-values from
Eq.~(\ref{eq:hydrodynamic-operator-transformation}). We expand the full
hamiltonian in $\delta \hat{a}_i$, keeping  terms up to quadratic order. The
first order term always vanishes as we are expanding about a minimum while the
zeroth order term gives the degenerate mean-field energies.  Thus we focus on
the second order contribution.

\begin{figure}
	\includegraphics{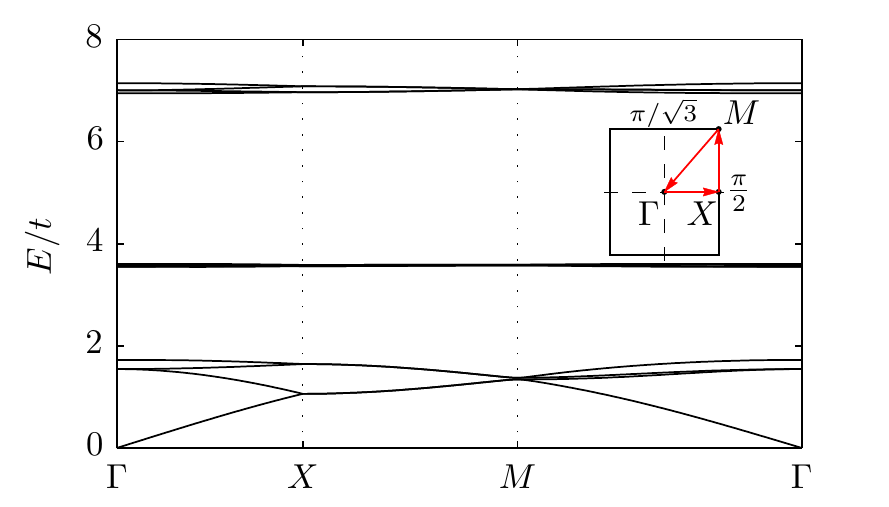}
	\caption{The twelve Bogoliubov modes about ground state (b) from
		Fig.~\ref{fig:korsh} at $U_*=U_\Delta$, $U_*/t=.5$ and
		$n_{*}=6$. As the interaction strengths $U_{*,\Delta}$
		decrease, the bands flatten  and the gaps between them approach
		$\sqrt{6} t$. At $U_{*,\Delta}=0$ we recover the dispersionless
		degenerate single-particle spectrum.}
	\label{fig:bogoliubov-spectrum-plot}
\end{figure}

Substituting $\hat{a}_i=a_i+\delta \hat{a}_i$ into
Eq.~(\ref{eq:full-hamiltonian}), and using the chemical potential given in
Eq.~(\ref{eq:mu}), one finds the quadratic Hamiltonian
\begin{align}
	\delta\hat{H}&=-t\sum_{\left\langle i j \right\rangle} (\ee^{\ii A_{i
	j}}\delta\hat{a}_i^\dagger\delta\hat{a}_j+ {\rm H.c.}) + \sum_i \left(
	U_i n_i+G_i \right)\delta\hat{a}_i^\dagger \delta\hat{a}_i
	\notag \\
	&+ \sum_i \frac{U_i }{2}\left( a_i^* a_i^*
	\delta\hat{a}_i\delta\hat{a}_i+a_i a_i
	\delta\hat{a}_i^\dagger\delta\hat{a}_i^\dagger \right)
\end{align}
where $n_i=\left| a_i \right|^2$, $G_*=2t\sqrt{\frac{3 n_\Delta}{n_*}}$, and
$G_\Delta=t\sqrt{\frac{3n_*}{n_\Delta}}$.  It greatly simplifies the analysis
to perform the gauge transformation $\delta\hat{a}_i \rightarrow
e^{i\theta_i}\delta\hat{a}_i$ at this stage.  This results in the following
\emph{gauge invariant} Bogoliubov Hamiltonian
\begin{align}
	\hat{H}_B&=-t\sum_{\left\langle i j \right\rangle} \left(\ee^{-\ii
	\Phi_{i j}}\delta\hat{a}_i^\dagger\delta\hat{a}_j+ {\rm H.c.}\right)
	\label{eq:awesome-bogoliubov-hamiltonian}
	\\
	&+ \sum_i \left[ \left( U_i n_i+G_i \right)\delta\hat{a}_i^\dagger
	\delta\hat{a}_i+\frac{U_i n_i }{2}\left(
	\delta\hat{a}_i\delta\hat{a}_i+
	\delta\hat{a}_i^\dagger\delta\hat{a}_i^\dagger \right) \right].
	\notag
\end{align}
The gauge invariant phase differences $\Phi_{ij}$  here are precisely those
introduced in Sec.~\ref{subsec:hydro}. 

We can again define momentum space operators $\delta\hat{a}_{\vec{k}\gamma} =
\frac{1}{\sqrt{N}} \sum_{n}\delta\hat{a}_{n \gamma}\ee^{-\ii
\vec{k}\cdot\vec{R}_n}$ with $\gamma=1,\cdots, M$, where $M$ is the number of
sites per unit cell. By expressing the Hamiltonian in terms of these operators
we obtain, up to a constant energy shift (equal for all mean-field states), the
Hamiltonian in the form
\begin{equation}
	\hat{H}_B =	\sum_{\vec{k}}
	\vec{\delta}\hat{\vec{a}}_{\vec{k}}^\dagger H_B(\vec{k})
	\vec{\delta}\hat{\vec{a}}_{\vec{k}}
	\label{eq:delta-Hamiltionian}
\end{equation}
where 
$\vec{\delta}\hat{\vec{a}}_{\vec{k}}=\left[ \delta\hat{a}_{\vec{k}1},
\cdots, \delta\hat{a}_{\vec{k}M},   
\delta\hat{a}_{-\vec{k}1}^\dagger, \cdots,
	\delta\hat{a}_{-\vec{k}M}^\dagger\right]^\top $ and 
\begin{equation}
  H_B(\vec{k})=
	\left[
		\begin{array}{c|c}
			C_{\vec{k}} & D \\
			\hline
			D & \raisebox{0pt}[1.2\height]{$C_{\vec{-k}}^\top$} 
		\end{array}
	\right].
  \label{eq:bogoliubov-matrix}
\end{equation}
Here $D$ is a diagonal matrix with $U_* n_*$  ($U_\Delta n_\Delta$) entries for
hub (rim) sites.  $C_{\vec{k}}=H_0(\vec{k})+G+D$, where $H_0(\vec{k})$ is the
single-particle Hamiltonian matrix appearing in
Eq.~(\ref{eq:momentum-space-Hamiltonian}), rewritten in the current gauge, and
$G$ is a diagonal matrix containing the values $G_*$ and $G_\Delta$ for  hub
and rim sites, respectively.

The creation and annihilation operators of the quasiparticle eigenstates of
this quadratic Hamiltonian will in general be a sum of both particle
annihilation and creation operators. These can be obtained through solving the
Bogoliubov de-Gennes (BdG) equations
\begin{equation}
	\eta H_B(\vec{k}) \vec{\phi}_{\vec{k}\gamma \pm} = \pm
	E_{\vec{k}\gamma} \vec{\phi}_{\vec{k}\gamma \pm}
\label{eq:BdG}
\end{equation}
where $E_{\vec{k}\gamma}\ge0$ and
$
\eta=\left( \begin{smallmatrix}\mathbb{1}_{M\times M} & 0 \\ 0 & -\mathbb{1}_{M
	\times M}\end{smallmatrix}\right)
$
(see, for instance, \cite{Blaizot1986}).  The energies of the Bogoliubov modes
are given by $ E_{\vec{k}\gamma}$ where $\gamma$ labels the band index.  The
quasiparticle operators which diagonalize $\hat{H}_B$ are determined from the
BdG eigenvectors as
$
\hat{\alpha}_{\vec{k}\gamma} = \vec{\phi}_{\vec{k}\gamma +}^\dagger \eta
\vec{\delta}\hat{\vec{a}}_{\vec{k}}
$.

The excitation spectrum  for a typical parameter set is shown in
Fig.~\ref{fig:bogoliubov-spectrum-plot}.  It is seen that the interactions give
dispersion to the excitation spectrum, which is completely flat at the
single-particle level.  The excitations about each vortex configuration  yield
a gapless Goldstone mode due to the broken U(1) superfluid phase.  These have
the dispersion $\sim \hbar \sqrt{(c_1 k_1)^2+(c_2 k_2)^2}$ where $k_{1,2} =
\vec{k} \cdot {\bf v}_{1,2}$ and $c_{1,2}$ is the speed of sound along the
${\bf v}_{1,2}$ lattice vectors.

\subsection{Computation of Degeneracy Lifting}

Having the excitation spectra at hand, we now move on  to discuss the resulting
degeneracy lifting.  We have calculated the thermal and quantum contributions
to the free energy in Eq.~(\ref{eq:free-energy}) at a range of values of the
input parameters $U_*/t$, $U_*/U_\Delta$, $n_*$ and, for the thermal part,
$T/t$. For each parameter configuration we obtained the band energies by
diagonalizing $H_B(\vec{k})$ from Eq.~(\ref{eq:bogoliubov-matrix}) at a
uniformly spaced grid of momenta in the Brillouin zone ~\cite{Monkhorst1976}.
Convergence as a function of the grid spacing was checked for each parameter
set.

\begin{figure}
	\includegraphics{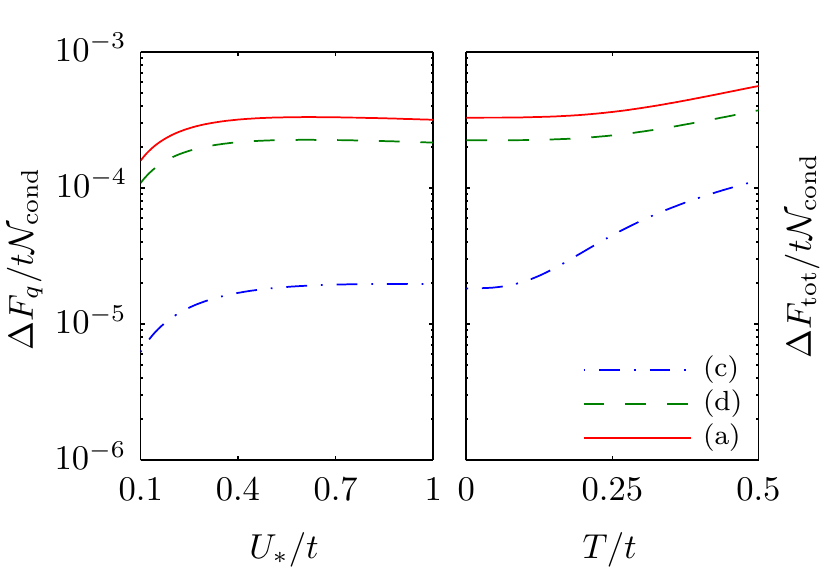}
	\caption{(Color online) Left: Quantum free energy difference per condensed particle
	with respect to state (b) from Fig.~\ref{fig:korsh} with $n_*=6,
	U_*=U_\Delta$. Right: Total free energy difference at the same $n_*$
	and $U_*/U_\Delta$, for finite temperature and $U_*/t=0.5$.}
	\label{fig:korsh-state-order}
\end{figure}

Results for a range of parameters are shown in
Fig.~\ref{fig:korsh-state-order}. We have plotted the differences of free
energies of states (a), (c) and (d) with respect to state (b),  $\Delta
F_{a,c,d} = F_{a,c,d}-F_b$ using the labelling of Fig.~\ref{fig:korsh}.  As seen
in the left-hand side of this figure, the resulting free energy difference is
always positive and so state (b) has the lowest free energy.   Thermal
fluctuations further enhance this degeneracy lifting as shown in the right-hand
side of this figure.

In addition to determining the ground state, we also observe that state (c) is
universally the highest in free energy. States (a) and (d) are typically
ordered as in Fig.~\ref{fig:korsh-state-order} but cases were found in which
their free energy curves cross.  The geometric mean of the sound speeds along
the two lattice vectors $\sqrt{c_1 c_2}$ is always lowest for (b) and highest
for  (c) which explains the ordering of the thermal contribution to the free
energy at low temperatures.

\subsection{The Condensate Depletion}
\label{sec:depletion}
Having established that state (b) has the lowest overall free energy, we now
move on to discuss its stability.  For Bogoliubov theory to be valid, one must
have that the number of particles excited out of the condensate is small
compared to the number of condensed particles.  The depletion, like the free
energy, can be separated into a quantum and thermal contribution, which we
denote by ${\cal N}_{\rm q}$ and  ${\cal N}_{\rm th}$, respectively. For the
above analysis to be correct we must have  ${\cal N}_{\rm dep}={\cal N}_{\rm
q}+{\cal N}_{\rm th}\ll  {\cal N_{\rm cond}}$.  From the solution of the
Bogoliubov-de Gennes equation (\ref{eq:BdG}) the depletion can be expressed as
\begin{align}
{\cal N}_{\rm q}&=\frac{1}{2}\sum_{\vec{k} \gamma}
\vec{\phi}^\dagger_{\vec{k}\gamma+} (\mathbb{1}-\eta)
\vec{\phi}_{\vec{k}\gamma+}
\\
{\cal N}_{\rm th}&=\sum_{\vec{k} \gamma} \vec{\phi}^\dagger_{\vec{k}\gamma+}
\vec{\phi}_{\vec{k}\gamma+}  f(E_{\vec{k}\gamma})
\end{align}
where $f(x)=\left( \ee^{\beta x} - 1 \right)^{-1}$ is the Bose Einstein
distribution function.

\begin{figure}
	\includegraphics{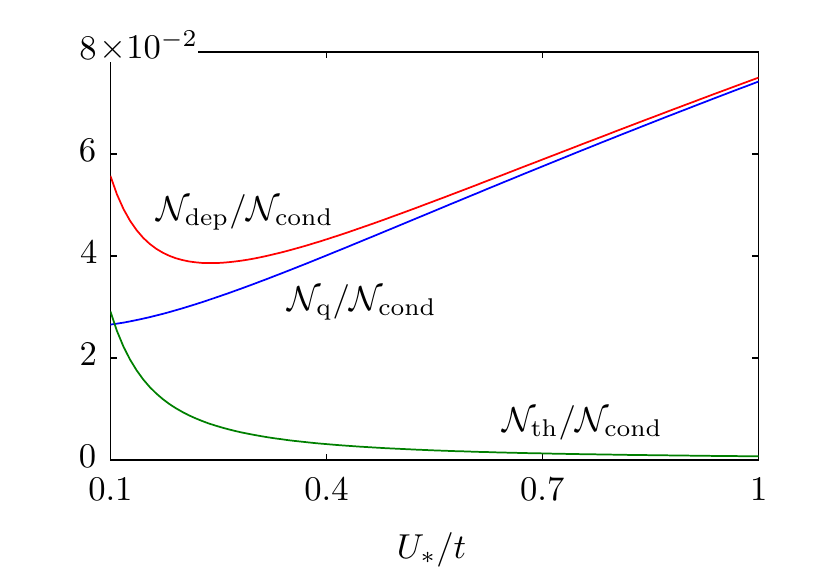}
	\caption{(Color online) The quantum, thermal and total depletion per condensed
		particle for a system consisting of $20\times 20$ unit cells at
		$n_*=6, U_*=U_\Delta$ and $T=t/10$.}
	\label{fig:depletion}
\end{figure}

While the quantum depletion converges, the thermal depletion integral has a
logarithmic infrared divergence due to  the Goldstone mode.  Such divergences
are typical for two-dimensional systems \cite{Mermin1966}.   Finite size
effects will remove this divergence and can be crudely taken into account  by
using a small-momentum cutoff of $2\pi/L$ where $L^2$ is the system size.
Consequently, the thermal depletion will scale as $\ln( L)$ for sufficiently
large $L$.

Figure~\ref{fig:depletion} shows the quantum and thermal contributions to the
total depletion at experimentally feasible parameters.  Quite interestingly,
the total depletion exhibits a non-monotonic behavior as a function of the
Hubbard interaction parameters.  In typical condensed systems, depletion
increases monotonically as a function of the interactions \cite{Pethick2008}.
A similar minimum was found for all parameter ranges tested. This can be
attributed to the flatness of the non-interacting band structure.  That is, as
interactions are decreased, the Bogoliubov band structure (c.f.
Fig.~\ref{fig:bogoliubov-spectrum-plot}) becomes flatter and so thermal
excitations are created more easily.  When $U_*=U_\Delta=0$ the Bogoliubov
spectrum reduces to completely flat bands and the thermal depletion will
diverge.  For the chosen parameters in this figure the depletion is always less
than 10\%.  The depletion can be further decreased by choosing larger average
density per site.

\section{Conclusion}
\label{sec:discussion}
In this work we have argued that, despite the large degeneracies associated
with the non-interacting spectrum of the dice lattice at half-flux per
plaquette, fluctuations will select a unique ground state when weak
interactions are included.  In particular, we have established that quantum and
thermal fluctuations select vortex lattice state (b) out of the  periodic
states shown in Fig.~\ref{fig:korsh} which are degenerate at the level of mean
field theory.  The stability of the resulting state was established for a range
of parameters by analyzing the quantum and thermal depletion.  

Although a thorough analysis of the non-periodic states is beyond the scope of
the current work, we will briefly comment on them now.  As discussed in
Sec.~\ref{sec:mft}, one can obtain vortex lattices (b,c,d) from state (a)
through the insertion of domain walls.  At the level of mean field theory,
these domain walls cost no energy to create and do not interact.  Therefore,
through insertion of domain walls one can obtain non-periodic vortex
configurations as well as vortex lattices with larger unit cell size, all of
which are degenerate at the level of mean field theory with the configurations
shown in Fig.~\ref{fig:korsh}.  On the other hand, including quantum and
thermal fluctuations will cause the domain walls to interact.  Preliminary
calculations of free energy shifts have in fact shown that type II domain walls
repel each other, so that the free energy is minimized when they are infinitely
separated which is equivalent to no domain walls being present.  On the other
hand, type I domain walls were found to attract each other, until state (b) is
reached.  This lends credence to the proposition that state (b) is indeed the
state of lowest free energy that is ultimately selected by quantum
fluctuations.  Considering this, a complete analysis of the
fluctuation-mediated interactions between the domain walls and their ensuing
dynamics should be an interesting direction for future work.

As has been noted near the end of section~\ref{subsec:ground-states},
macroscopic considerations may also allow for more energetic mean-field domain
walls and point defects. Another interesting direction for future research would be
to verify whether such defects are indeed stable. Consider again the double
asymptotically domain wall-free configuration described at the end of
Sec.~\ref{subsec:ground-states}. From the viewpoint of the time-dependent
Gross-Pitaevskii equation, we can imagine assigning random vorticities and
consistent phases, if they can be found, to the plaquettes in the intermediate
region and random particle numbers to the sites and then tracking their time
evolution.  Interestingly, the plaquette vorticities cannot change through
smooth time evolution, so we should expect different vorticity distributions to
correspond to distinct hydrodynamic solutions. However, there might not always
be global steady-state solutions. If, on the other hand, energetic domain walls
and point defects are found to be stable, it would enable numerous further
investigations to be carried out, such as of the interplay between massive and
zero-energy domain walls and their free energy-mediated interactions.

\acknowledgements
We would like to thank Dmitry Kovrizhin and Gunnar M\"{o}ller for
fruitful discussions. We are grateful for funding from a Roth
Studentship from the Imperial College London Department of Mathematics
(MP) and the European Union's Seventh Framework Programme for
research, technological development, and demonstration under grant
agreement PCIG-GA-2013-631002 (RB).

\bibliography{sources}
\end{document}